\documentclass[twocolumn,aps,prb,superscriptaddress]{revtex4}
\usepackage{graphics,bm,graphicx}
\usepackage{amssymb}
\usepackage{epsfig}
\usepackage{epsf}

\def\be{\begin{equation}} \def\ee{\end{equation}}
\def\beq{\begin{eqnarray}} \def\eeq{\end{eqnarray}}

\def\nn{\nonumber}

\begin{document}
	
	\title{Mechanism of the resistivity switching induced by the Joule heating in crystalline NbO$_2$}
	
	\author{Samuel W. Olin}
	\email{solin1@binghamton.edu}
	\affiliation{Department of Physics, Applied Physics, and Astronomy, Binghamton University, Binghamton, New York, 13902, USA}
	
	\author{S. Abdel Razek}
	\affiliation{Department of Physics, Applied Physics, and Astronomy, Binghamton University, Binghamton, New York, 13902, USA}
	
	\author{L. F. J. Piper}
	\affiliation{WMG, University of Warwick, Coventry CV4 7AL, UK}
	
	\author{Wei-Cheng Lee}
	\email{wlee@binghamton.edu}
	\affiliation{Department of Physics, Applied Physics, and Astronomy, Binghamton University, Binghamton, New York, 13902, USA}

	\date{\today}
	
	\begin{abstract}
		Recently the memristive electrical transport properties in NbO$_2$ have attracted much attention for their promising application to the neuromorphic computation. At the center of debates is whether the metal-to-insulator transition (MIT) originates from the structural distortion (Peierls) or the electron correlation (Mott).
		With inputs from experiments and first principles calculations, we develop a thermodynamical model rooted in the scenario of the MIT driven by a $2^{nd}$ order Peierls instability. We find that the temperature dependence of the electrical conductivity can be accurately fit by the band gap varying with temperature due to the gradual weakening of the Nb-Nb dimers. The resistivity switching can consequently be understood by dimer-free metallic domains induced by  local Joule heating. In solving the heat equation, we find that the steady state can not be reached if the applied voltage exceeds a threshold, resulting in the chaotic behavior observed in the high voltage and current states. With the Ginzburg-Landau theory and the Joule heating equation, the evolution of the metallic domains under bias voltage can be simulated and directly verified by experiments.
	\end{abstract}
	
	\maketitle
	
	\section{Introduction}
	Materials exhibiting resistivity switching under the bias voltage have regained enormous interest in the past decade due to their potential applications to neuromorphic computation.\cite{yang2013,yu2018} After the discovery of the nanoscale memristor in TiO$_x$,\cite{strukov2008} great progress has been made to investigate oxide-based memristors.\cite{wang2015,xu2021} For typical binary oxides, including TiO$_2$, TaO$_2$, ZrO$_2$, HfO$_2$, etc.\cite{strukov2008,abbas2018,choi2018,goodwill2019,yan2019,pahinkar2020,athena2022} the resistivity switching is generally attributed to conductive filaments formed by mobile oxygen vacancies.
	Recently, correlated oxides VO$_2$\cite{kumar2013,delValle2019} and NbO$_2$\cite{pickett2013,kumar2017-na,kumar2017} have come to the spotlight in the study of memristors because their resistivity switching can not be explained by the filament mechanism. These materials have an intrinsic metal-to-insulator transition (MIT) in the electronic structures at a critical temperature $T_c$. As a result, if the Joule heating effect is large enough to induce the MIT {\it locally}, metallic domains can be formed to provide conducting pathways for electrons, leading to a novel mechanism of the resistivity switching. Since the MIT is related to the electron dynamics driven mainly by temperature and no ionic motions are needed, this MIT-based mechanism provides unique advantages of rapid switching speed and low energy cost.

	The fundamental issue that must be addressed is identifying the driving force of the MIT. One possibility is the electron correlation, namely {\it Mott physics}, while another is the spontaneous dimerization due to the structural distortion, namely the {\it Peierls instability}.  
	It is widely recognized that Mott physics is important in VO$_2$, despite abundant evidence strongly suggesting that the structural change is not negligible either.\cite{eyert2002,haverkort2005,lazarovits2010,brito2016,mukherjee2016,paez2020,evlyukhin2020,mondal2021,singh2022} The debate for NbO$_2$, on the other hand, remains unresolved. While the density-functional theory (DFT) predicts an energy gap opening in the band structure in the presence of the Nb-Nb dimers without the Mott physics involved, indicating the nature of MIT to be the Peierls transition,\cite{ohara2014,ohara2015,wahila2019} the DFT implemented with the dynamical mean-field theory (DFT+DMFT) suggests that Mott physics is necessary to predict the correct value of the gap.\cite{brito2017} From an experimental perspective, the hard X-ray photoemission spectroscopy (HAXPES) shows no hint of the lower Hubbard band, a hallmark signature of Mott physics, \cite{leewc2019,wahila2019} and the gradual weakening of the Nb-Nb dimer with the increase of the temperature has been clearly observed in the temperature-dependent extended X-ray absorption fine structure spectroscopy (T-EXAFS).\cite{galo2021}
	Due to the recent rapid growth of interest in memristor devices based on NbO$_2$\cite{pickett2013,kumar2017-na,kumar2017}, resolving the nature of MIT in NbO$_2$ becomes a crucial task in order to have a proper model to simulate the performance of NbO$_2$ devices.\cite{nandi2015,funck2016,kumar2017-na,kumar2017,messaris2020}
	
	Models based on Poole-Frenkel conduction have been proposed to describe the resistance change in NbO$_x$ devices.\cite{kumar2017-na,kumar2017,leepssr}  These models are successful in describing novel resistivity changes at low temperatures due to non-linear electrical transport induced by extrinsic effects like defects, trapped potential, dimension reduction in the thin films, and so on.
	Contrastingly, in this paper we present a thermodynamical model for understanding and simulating the resistivity switching of the NbO$_2$ as an active layer in memristor devices resulted from the intrinsic quantum mechanical property of the MIT at higher temperatures. Based on the scenario of the second order Peierls transition as the MIT mechanism, we construct a classical Monte Carlo (MC) simulation for the Ginzburg Landau theory that can accurately reproduce the temperature evolution of the Nb-Nb dimer length measured by T-EXAFS. Using a temperature-varying band gap due to the gradual weakening of Nb dimers, we show that the electrical conductivity can also be accurately reproduced within a very wide range of temperature  ($1\leq10^3/T\leq 6$).
	
	Combining the Ginzburg-Landau theory with the Joule heating equation, we are able to directly simulate the evolution of the metallic domains with a bias voltage. We will show that because of the highly non-linear temperature dependence of the electrical conductivity, there exists a threshold voltage above which no steady state solution will be achieved in the insulating phase. We argue that this absence of the steady state solution in the insulating phase gives an natural explanation of the chaotic behavior observed in the high voltage and current states. We will further 
	demonstrate that the threshold voltage could be experimentally engineered in the system having the geometry of a three-dimensional cylinder with axial symmetry. Based on our results, we propose that the metallic domains with the weakened Nb-Nb dimer formed by the Joule heating effect is the main mechanism of the resistivity switching in NbO$_2$, and our prediction can be further tested by the T-EXAFS measurement on the NbO$_2$ under bias voltage. Our work opens a new route to achieve the voltage-induced resistivity switching in a crystalline insulator with a second order Peierls transition.
	
	\section{Models}
	\subsection{Ginzburg-Landau theory}
	\begin{figure}
		\includegraphics[width=3.5in]{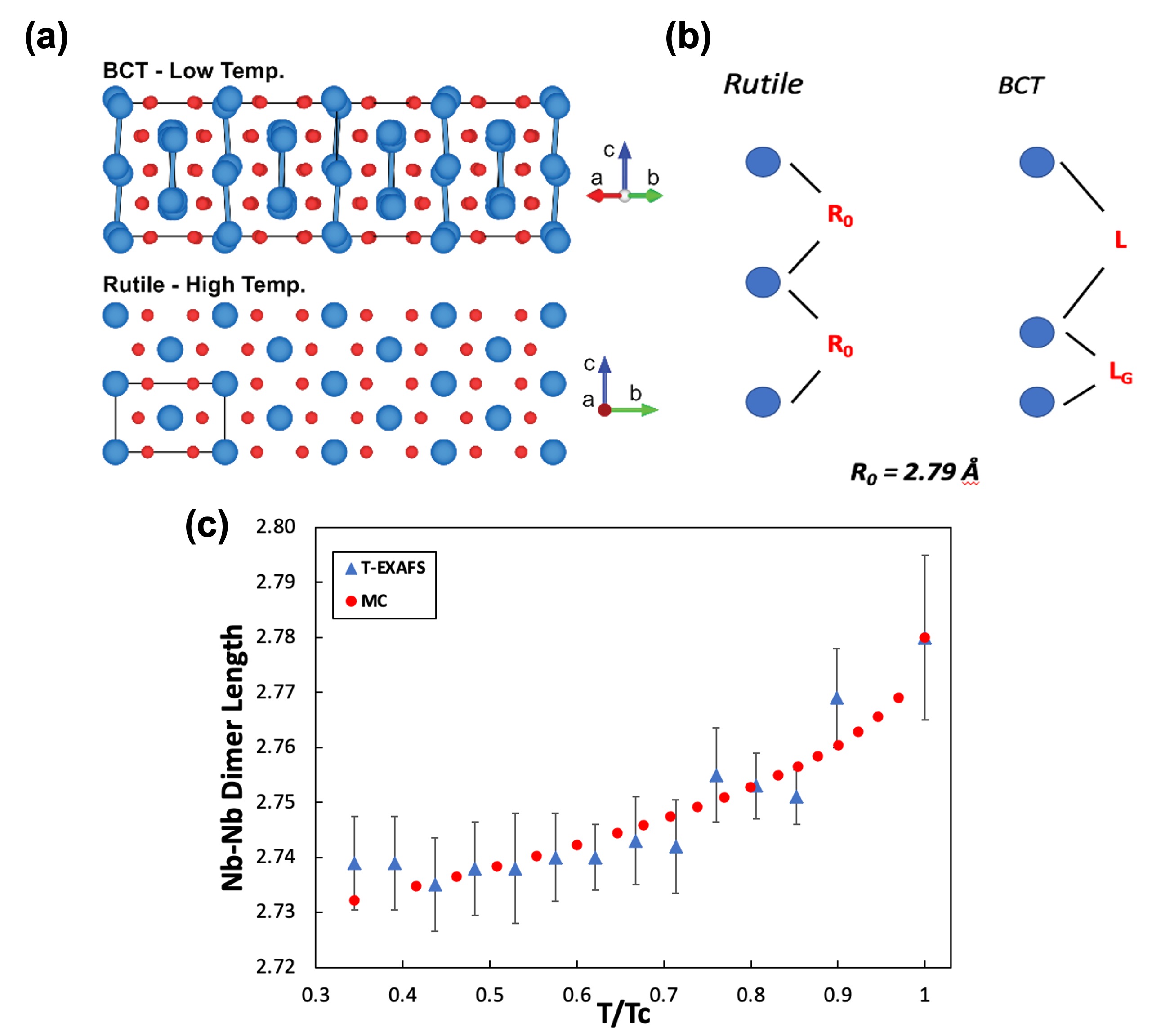}
		\caption{\label{fig:order} (a) The crystal structures of NbO$_2$ in metallic (rutile) phase and in the insulating body-centered-tetragonal (BCT) phase. This picture is reproduced from Ref. [\onlinecite{wahila2019}]. (b) The schematical illustration of the definition for the order parameter. $R_0$ is the distance between nearest neighbor Nb atoms in the metallic rutile (undimerized) phase. $L$ and $L_G$ are the longer and the shorter Nb-Nb distances in the insulating BCT (dimerized) phase, and they are subject to a constraint of $L+L_G = 2R_0$. The order parameter for the second order Peierls transition is chosen to be $M= \frac{L-R_0}{R_0} = \frac{R_0-L_G}{R_0}$. Note that $L_G$ is the Nb-Nb dimer length directly measured by T-EXAFS in previous work.\cite{galo2021} (c) The comparison of our Monte Carlo simulation on the Ginzburg Landau theory for the Nb-Nb dimer length ($L_G$) as a function of temperature with the experimental data. The error bars are for the experimental data.}
	\end{figure}

	The thermodynamical properties related to a second order phase transition can be well described by the 
	Ginzburg-Landau (GL) free energy in the general form of
	\be
	F_{GL} = F_0 + \frac{a(T)}{2}M^2 + \frac{b(T)}{4}M^4 + \frac{c(T)}{6}M^6,
	\ee
	where $a(T)$, $b(T)$, and $c(T)$ are some functions of the temperature $T$. $F_0$ is the free energy in the normal state without the order, and it is typically irrelevant to the dynamics of the order parameter we want to investigate. $M$ is the order parameter characterizing the phase transition. In the case of the second order Peierls transition, $M$ is naturally chosen to be a dimensionless quantity related to the Nb-Nb dimer length as
	\be
	M= \frac{L-R_0}{R_0} = \frac{R_0-L_G}{R_0},
	\ee
	where $R_0$ is the distance between nearest neighbor Nb atoms in the metallic rutile (undimerized) phase. $L$ and $L_G$ are the longer and the shorter Nb-Nb distances in the insulating body-centered-tetragonal (BCT), dimerized phase, and they are subject to a constraint of $L+L_G = 2R_0$.  The graphic representation of the order parameter can be found in Fig. \ref{fig:order}(a) and (b).
	
	Our goal is to use the standard approach of the GL theory to fit the T-EXAFS data\cite{galo2021}, and we perform the Monte Carlo simulation with the GL free energy to compute the Nb-Nb dimer length. The details of the MC simulation can be found in Appendix \ref{amc}, and the result is presented in Fig. \ref{fig:order} (c). 
	Our simulation reproduces the experimental data accurately within the experimental error bars, which is a strong support of the Peierls transition in NbO$_2$ being second order in nature.
	
	The GL theory can be generalized to the inhomogeneous case in which the order parameter is allowed to vary spatially. This generalization is necessary as we consider the effect of inhomogeneous temperature distribution $T(\vec{r})$ induced by the Joule heating. This could result in the inhomogeneous order parameter even in a ideal crystalline structure, leading to the formation of metallic domains responsible for the novel resistivity switching observed in experiments. The simulation of metallic domains will be presented in Sec. \ref{sec:gld}.
	
	\subsection{Joule heating equation}
	To obtain the temperature distribution $T(\vec{r})$, we solve the Joule heating equation derived from the general continuity equation of the heat energy flow. As the electrical current flows through the system, electrons can transfer energy into the system by interacting with ions via electron-phonon couplings, which generate heat locally. On the other hand, the heat energy generated locally could be dissipated through the thermal conductivity. As a result, the conservation of energy requires
	\be
	\rho_m C\frac{\partial T}{\partial t} = \sigma(T(\vec{r})) \vert-\vec{\nabla} \phi\vert^2 + \kappa(T(\vec{r}))\nabla^2 T,
	\label{jhfull}
	\ee
	where $\rho_m$, $C$, $\sigma(T)$, and $\kappa(T)$ are the mass density, the heat capacity, the electrical conductivity, and the thermal conductivity of the system.\cite{nandi2015,basnet2020} $\phi$ is the electrical potential due to the application of the external voltage $v_0$. In order to have a Joule heating equation that could descibe realistic NbO$_2$ devices, we need accurate models for both the electrical conductivity and thermal conductivity.
	
	\subsubsection{Electrical conductivity}
	
	\begin{figure}
		\includegraphics[width=3in]{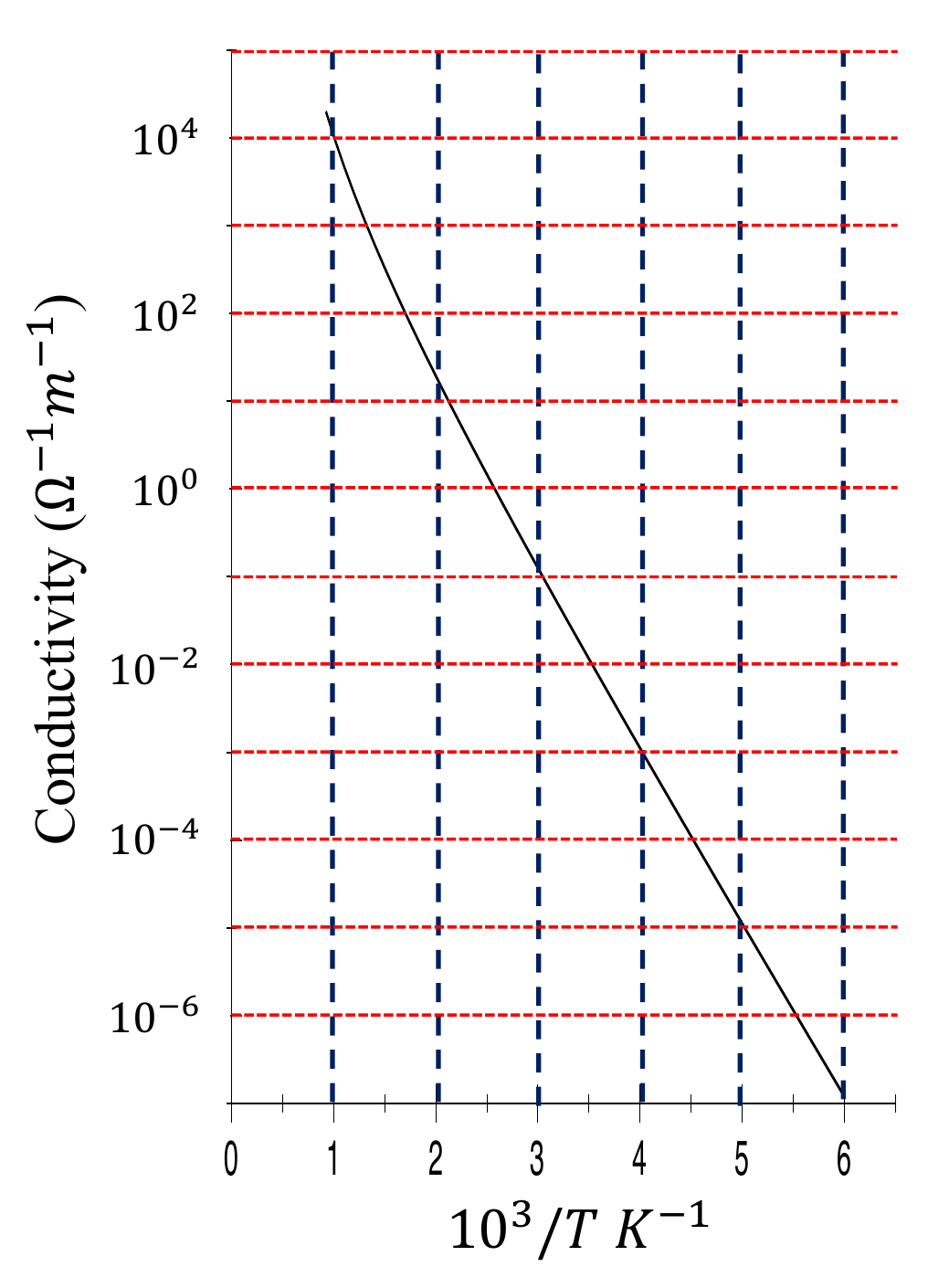}
		\caption{\label{fig:sigma} The electrical conductivity of NbO$_2$ obtained by the proposed fit. This result may be verified directly with experiment, and is in an excellent agreement with the measurement performed by Sakai {\it et. al.}\cite{sakai1985}}
	\end{figure}
	
	It has been observed that the electrical conductivity of NbO$_2$ exhibits a temperature dependence that deviates from the typical behavior of a semiconductor. \cite{sakai1985} Following from our scenario of the second order Peierls transition, the gap becomes temperature-dependent due to the weakening of Nb-Nb dimerization, which can be described by the following universal function: $\Delta(T) = \Delta_0 - aT^2$. Since the gap should be closed at $T=T_c$, the parameter $a$ can be fixed by $a=\Delta_0/T^2_c$. Consequently, the final form of the gap as a function of $T$ becomes
	\be
	\Delta(T) = \Delta_0(1-\frac{T^2}{T_c^2}).
	\ee
	Using the above temperature-dependent gap function, we introduce the fitting function for the NbO$_2$ electrical conductivity as
	\be
	\sigma(T)=\sigma_0\exp\left[\beta \Delta_0(1-\frac{T^2}{T_c^2})\right],
	\ee
	where $\beta = 1/k_B T$. 
	
	Note that $\sigma_0$ should be the conductivity in the metallic phase ($\sigma(T=T_c)=\sigma_0$), which is known to be around $20000$ $\Omega^{-1}m^{-1}$. Moreover, the value of $T_c$ is experimentally determined to be $T_c=1083 K$. \cite{sakai1985} In other words, our proposed fitting function in fact has only one free parameter $\Delta_0$. The result of our fitting with $\Delta_0=0.38$ eV is plotted in Fig. \ref{fig:sigma}, which shows an excellent agreement with the experimental data. Furthermore, it is noted that $\Delta_0$ can be interpreted as the energy gap at $T=0$, and the value of $0.38$ eV is within the reasonable range obtained in the previous DFT calculations.\cite{ohara2015,wahila2019} The success of our fitting function to the electrical conductivity provides additional evidence for the second order Peierls transition in NbO$_2$ being the MIT mechanism in NbO$_2$.
	
	\subsubsection{Thermal conductivity}
	The thermal conductivity of NbO$_2$ has been a great interest in the community of thermoelectric materials for its promising application to the non-volatile resistive random-access memory (RRAM).\cite{nandi2015} However, the availability of experimental and theoretical data is very limited in the existing literature.
	Recently, the thermal conductivity of NbO$_2$ thin films with different orientations has been reported to be within the range of 2.5 to 3 W m$^{-1}$ K$^{-1}$ at room temperature.\cite{cho2019} It has also been found that the amorphous phases of NbO$_{2-x}$ could exhibit a thermal conductivity in a wide range from less than 1 W m$^{-1}$ K$^{-1}$ to 7 W m$^{-1}$ K$^{-1}$ due to the mixture of NbO$_x$ and Nb$_2$O$_{5-x}$. \cite{cheng2019,music2020}
	The DFT calculation of thermal conductivity is challenging as well due to the phonon soft modes that emerge from the nature of the second order Peierls transition.\cite{ohara2014,ohara2015}
	Nevertheless, since the phonon soft modes are highly damped, they do not contribute significantly to the thermal conductivity. As a result, it is expected that the main contribution is still from the acoustic phonons and the DFT calculation should still give a reasonable value of the thermal conductivity even in the presence of phonon soft modes. 
	
	Since the experimental data are available in the insulating but not in the metallic phases, we employ the DFT to calculate the thermal conductivity in the metallic rutile phase. The details of the DFT calculations can be found in Appendix \ref{adft}.
	Fig. \ref{fig:ther}(a) shows the phonon dispersion in the metallic rutile phase obtained by our DFT calculations, and we find the phonon soft modes around $R$ and $A$ points, consistent with Ref [\onlinecite{ohara2015}]. As shown in Fig. \ref{fig:ther}(b), the thermal conductivity is found to be 4.4 W m$^{-1}$ K$^{-1}$ at room temperature and decreases to 2.6 W m$^{-1}$ K$^{-1}$ at the transition temperature $T_c\approx 1083 K$, using the phonon dispersion obtained in 
	Fig. \ref{fig:ther}(a). We also check the Debye temperature $\theta_D$ inferred from the heat capacity (Fig. \ref{fig:ther}(inset)), and we obtain $\theta_D \approx 653K$, similar to the value reported in Ref [\onlinecite{ohara2015}].
	
	\begin{figure}
		\includegraphics[width=3in]{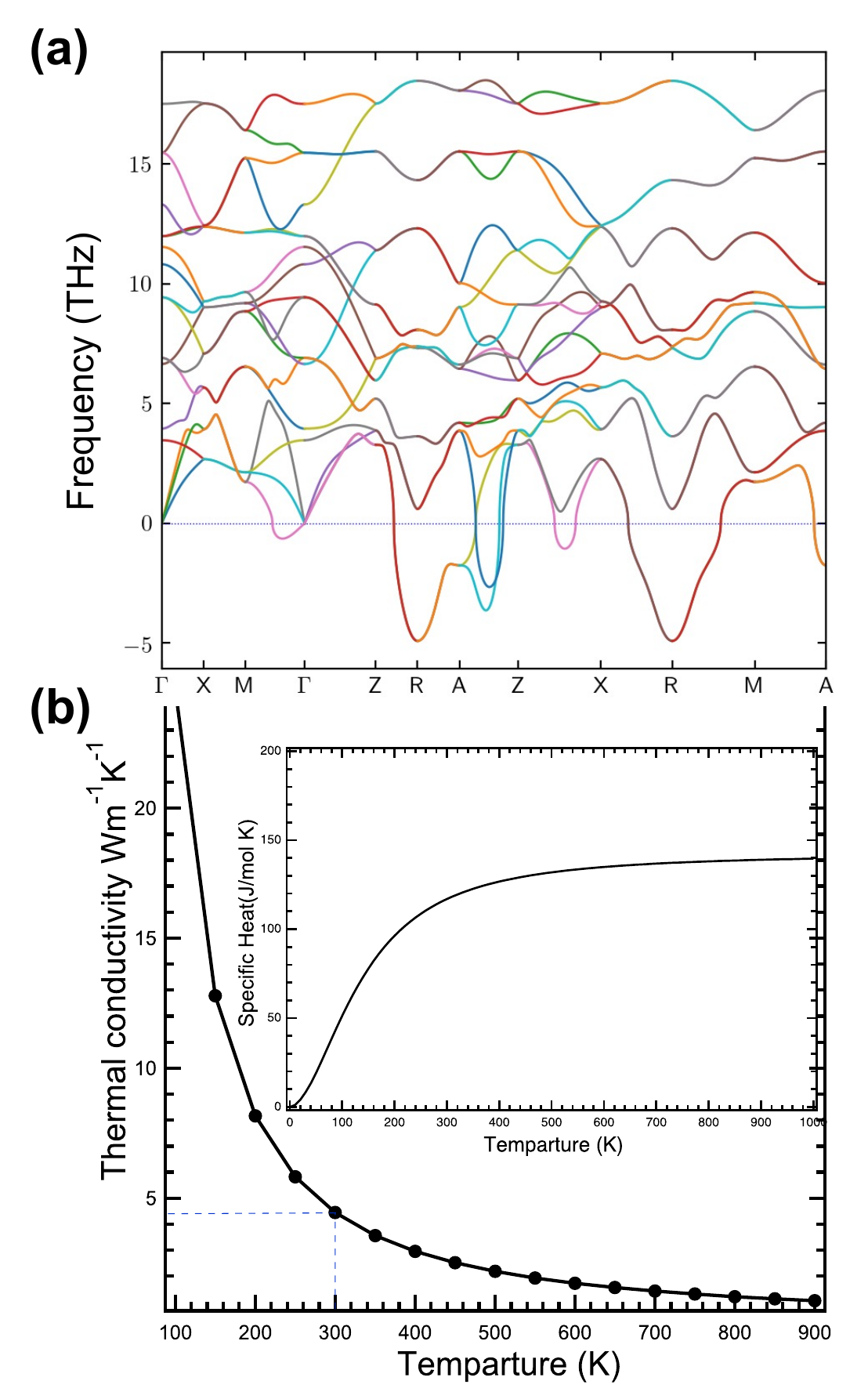}
		\caption{\label{fig:ther} (a) The phonon dispersion calculated by DFT for metallic rutile NbO$_2$. There are phonon soft modes corresponding to the instability toward the Nb-Nb dimerization. (b) The thermal conductivity in the metallic rutile NbO$_2$. {\it Inset}: The heat capacity as a function of the temperature indicates the Debye temperature to be around 653K.}
	\end{figure}
	
	Based on the discussion of the thermal conductivity given above, we introduce the thermal conductivity in the following form:
	\beq
	\kappa(T)&=& \kappa_{m}, T\geq T_c,\nn\\
	&=& \kappa_{i}, T<T_c.
	\eeq
	$\kappa_m$ and $\kappa_i$ represent the thermal conductivity in the metallic and insulating phases respectively, and their values are taken to be 
	$\kappa_{m} = 2.6$ W m$^{-1}$ K$^{-1}$ and 
	$\kappa_{i} = 0.156$ W m$^{-1}$ K$^{-1}$ respectively.
	Similar modeling of the thermal conductivity for NbO$_2$ devices has been used in a previous study.\cite{nandi2015}
	
	\section{Results}
	
	\subsection{Crossover from three dimensions to one dimension}
\begin{figure}
		\includegraphics[width=3.5in]{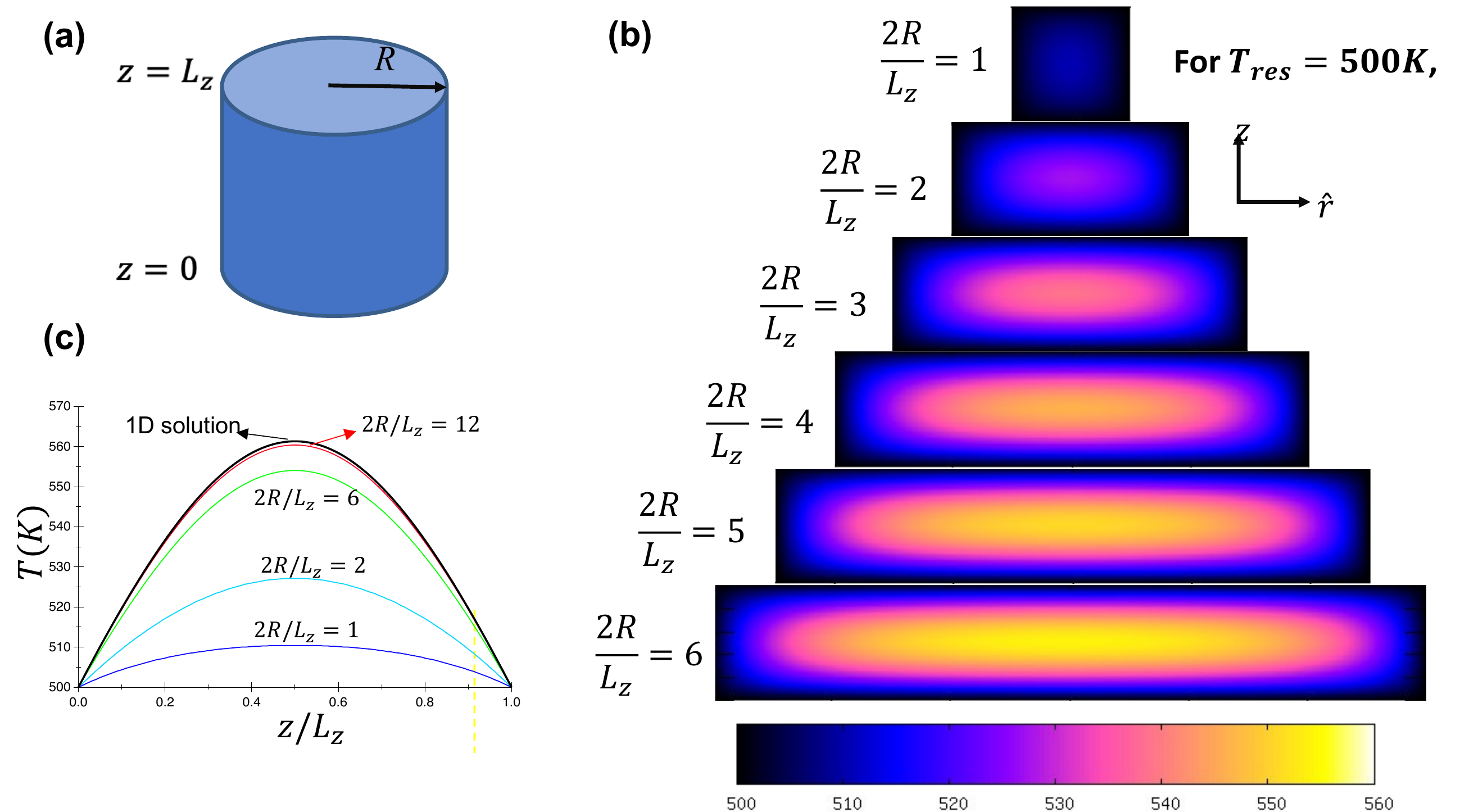}
		\caption{\label{fig:3d} (a) The geometry of the simulated system. The cylindical coordinate $(r,\theta,z)$ is used, and the temperature varies as functions of $(r,z)$ only. (b) The temperature distribution in the $(r,z)$ plane for the reservoir temperature $T_{res}=500K$ and the external voltage $v_0 = 1.2$ V as a function of the ratio of the diameter to longitudinal length $2R/L_z$. the maximal temperature at the center increases initially and then saturates to a value that depends only on $T_{res}$ and $v_0$. (c) At large $2R/L_z$ , the temperature distribution at the center ($(r=0)$) saturates to the 1D steady state solution.}
	\end{figure}

	Throughout this paper, we assume that the geometry of the system is a three-dimensional cylinder with axial symmetry as shown in Fig. \ref{fig:3d}(a) unless stated otherwise. As a result, we adopt the cylindrical coordinate $(r,\theta,z)$ and assume that all the physical properties depend only on $(r,z)$. We will focus on the steady state solution, thus any time-dependence will be ignored too. The Joule heating equation for the steady state solution can consequently be reduced to 
	\beq
	&&\sigma(T(r,z)) \vert-\vec{\nabla} \phi\vert^2 + \kappa(T(r,z))\nabla^2 T = 0,\nn\\
	&&\nabla^2 T = \left[\frac{1}{r}\frac{\partial}{\partial r}\left(r \frac{\partial T(r,z)}{\partial r}\right) + \frac{\partial^2 T}{\partial z^2}\right],
	\eeq
	where the derivative with respect to $\theta$ is dropped due to the axial symmetry. We employ the finite element method with the boundary condition of
	\be
	T(r,0) = T_1\,\,\,,\,\,\,T(r,L_z) = T_2\,\,\,,\,\,\,
	T(R,z) = T_3
	\ee
	for any $r$ and $z$. This boundary condition reflects the fact that the system is in contact with thermal reservoirs from the bottom electrode at $z=0$, top electrode at $z=L_z$, and lateral surface at $r=R$ fixed at temperatures of $T_1$, $T_2$, $T_3$ respectively. In principle, the Joule heating equation can be solved with any values of $(T_1,T_2,T_3)$. Since we are only interested in the temperature variation induced by the Joule heating effect inside the system and do not consider the application of external temperature gradient, we will mainly focus on the steady state solution with $T_1=T_2=T_3=T_{res}$, where $T_{res}$ is a fixed temperature reservoir. $\phi$ is the electrical potential giving rise to the electrical field of $\vec{E} = -\vec{\nabla} \phi$ that can be approximated to be $-v_0/L_z$, where $v_0$ is the voltage different between the top and the bottom electrodes.
	
	Fig. \ref{fig:3d}(b) plots the temperature distribution in the $(r,z)$ plane with $T_{res}=500K$ and $v_0=1.2$ V for different ratios of the diameter to longitudinal length $2R/L_z$. A $T_{res}$ of $500K$ was chosen because the electrical conductivity begins showing highly nonlinear behavior at this temperature and above, as seen in Fig. \ref{fig:sigma}. We find that as $2R/L_z$ increases, the maximal temperature at the center increases initially and then saturates to a value that depends only on $T_{res}$ and $v_0$.
	The physics of this saturated temperature can be understood as follows. In the steady state solution, the Joule heating effect pumps in energy locally via the term of $\sigma(T(r,z)) \vert-\vec{\nabla} \phi\vert^2$, and this energy has to be removed by the thermal conductivity via the term of $\kappa(T(r,z))\nabla^2 T$. Therefore, the only way to comprise both effects is to develop a temperature variation throughout the system. In the limit of large $2R/L_z$, the gradient of the temperature along the $\hat{r}$ direction near the center ($r=0$) is nearly zero, thus the heat energy can be dissipated only via the temperature variation along the $\hat{z}$ direction. In other words, the temperature distribution at the center is effectively described by the one-dimensional version of the Joule heating equation as
	\be
	\sigma(T(0,z)) \left(-\frac{\partial \phi}{\partial z}\right)^2 + \kappa(T(0,z))\frac{\partial^2 T}{\partial z^2} = 0.
	\label{jhe-1d}
	\ee
	Fig. \ref{fig:3d}(c) presents the temperature distribution at the center for different values of $2R/L_z$ together with the 1D solution, which confirms the asymptotical behavior of the crossover from three dimensions to one dimension discussed above.
	
	\subsection{Threshold voltage for the steady state solution}
	Another intriguing feature is the existence of a threshold voltage $v_c$ above which the steady state solution yields a temperature larger than $T_c$ {\it everywhere} excluding the boundaries. In other words, if the applied voltage exceeds $v_c$, no steady state solution exists in the insulating phase, and the entire system will be driven to the metallic phase. The existence of the threshold voltage results from the highly non-linear temperature dependence of the electrical conductivity $\sigma(T)$. As seen from the heat equation, the energy pumped into the system via the Joule heating is proportional to the electrical conductivity. In the temperature range where $\sigma(T)$ increases significantly with $T$, the temperature variation required to dissipate the heat energy will need to increase too. However, because $\sigma(T)$ is highly non-linear in $T$, the temperature will continue to grow drastically until $\sigma(T)$ no longer increases non-linearly with $T$, i.e., until the metallic phase is reached. 
	
\begin{figure}
		\includegraphics[width=3in]{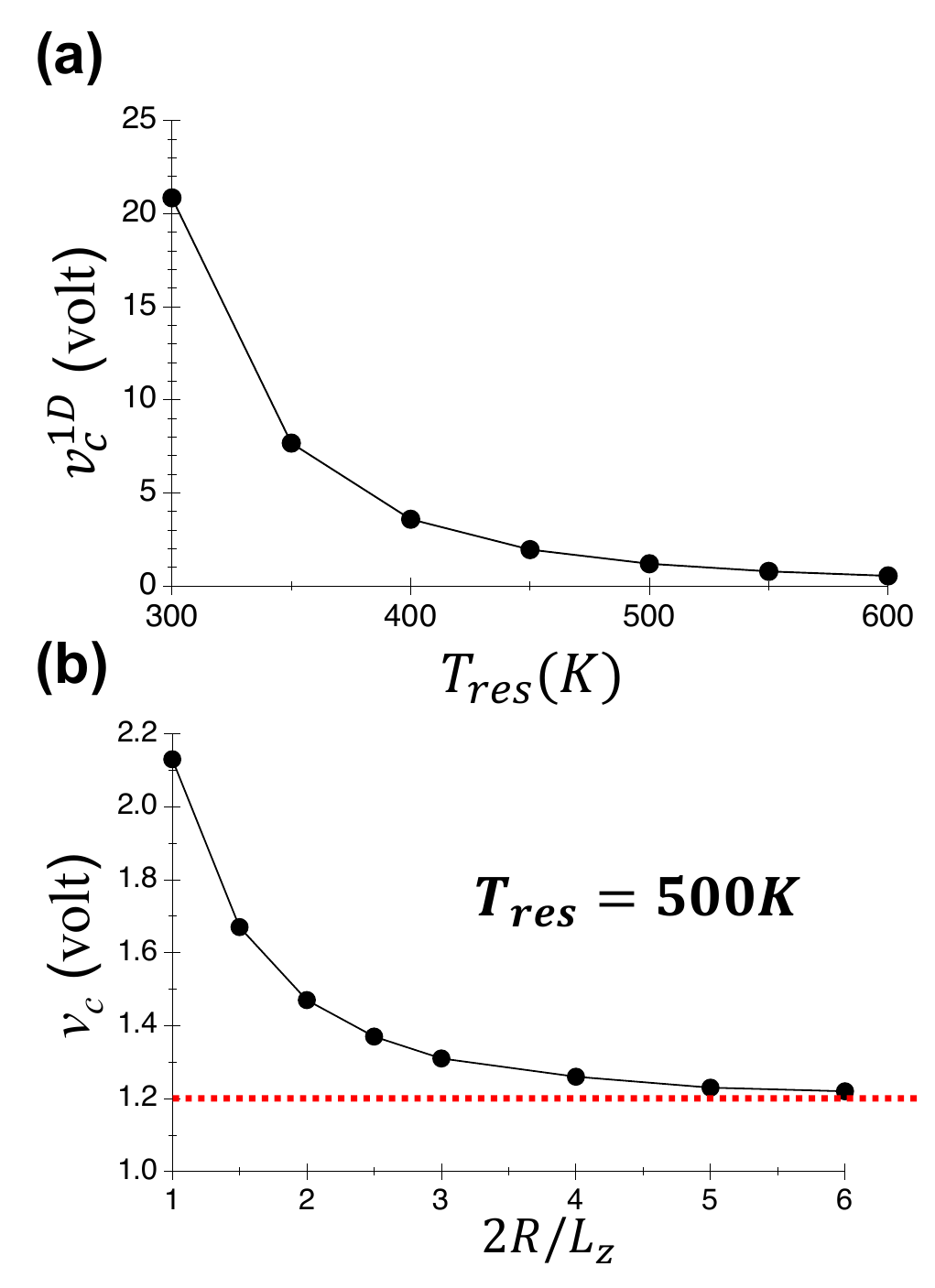}
		\caption{\label{fig:vc} (a) The threshold voltage obtained in the one-dimensional limit $v_c^{1D}$ as a function of the reservoir temperature $T_{res}$. (b) The threshold voltage $v_c$ in a 3D cylindrical geometry as a function of $2R/L_z$. Observe that $v_c$ decreases as $R/L_z$ increases and saturates to the value of $v_c$ obtained in the one-dimensional limit. }
	\end{figure}

	For the proof of concept, we analyze the threshold voltage in the one-dimensional limit ($v_c^{1D}$), and the result is presented in Fig. \ref{fig:vc}(a). In general, we find that the increase of $\sigma(T)$ becomes much larger in the range of $500K$ to $T_c$, and $v_c^{1D}$ is of the order of 1 volt or less in this regime.
	Now we come back to the three-dimensional case, and
	Fig. \ref{fig:vc}(b) shows $v_c$ as a function of $2R/L_z$ for $T_{res}=500 K$. We observe that $v_c$ decreases as $2R/L_z$ increases and saturates to the value of $v_c^{1D}$ obtained in the one-dimensional limit, confirming again the asymptotical behavior discussed above. 
	
	Our result suggests that the crossover from three dimensions to one dimension can be used as a parameter for tuning the performance of NbO$_2$ devices. It is emphasized that the crossover is not limited to the system with a cylindrical geometry. Instead, the crossover could generally occur if the length along the longitudinal direction differs significantly from the length along transverse directions. Practically,
	the threshold voltage found in the one-dimensional limit $v_c^{1D}$ will be the minimum of the threshold voltage for a 3D system, and engineering the ratio of $2R/L_z$ could be a unique way to tune the threshold voltage in realistic NbO$_2$ devices. The physical implication of the threshold voltage will be discussed in Sec. \ref{sec:diss}.
	
	\subsection{Metallic domain of Nb-Nb weakened dimers in the steady state}
	\label{sec:gld}
\begin{figure*}
		\includegraphics[width=7in]{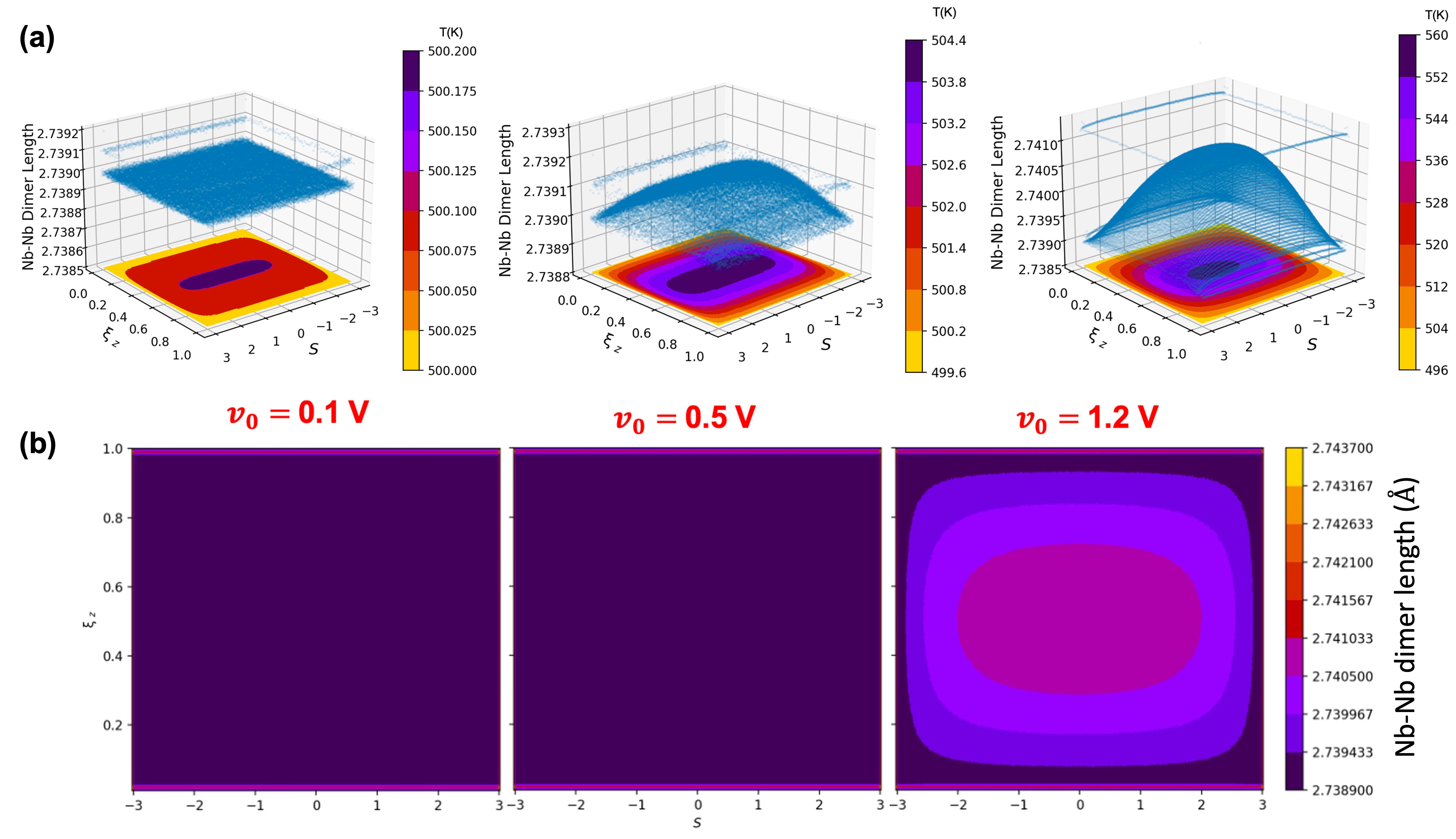}
		\caption{\label{fig:dimer} (a) The profiles of Nb-Nb dimer length $L_G$ in the $(r,z)$ plane for $v_0=0.1, 0.5, 1.2$ V at the reservoir temperature of $T_{res}=500 K$. The scattered plot on the top is for $L_G$, and the color contour at the bottom is for the temperature $T$. (b) The comparison of Nb-Nb dimer length $L_G$. The metallic domain is clearly seen for $v_0=1.2 V$ but not for $v_0=0.1 V$ and $v_0=0.5V$.}
	\end{figure*}

	In the presence of the temperature distribution induced by the Joule heating, the order parameter becomes inhomogeneous.
	The corresponding GL theory describing $M(\vec{r})$ could be written as
	\beq
	&&F_{GL} - F_0\nn\\
	&=& \int d\vec{r}\frac{a(T(\vec{r}))}{2}M^2(\vec{r}) + \frac{b(T(\vec{r}))}{4}M^4(\vec{r}) + \frac{c(T(\vec{r}))}{6}M^6(\vec{r})\nn\\
	&+& \gamma \int d\vec{r} \vert \vec{\nabla} M\vert^2,
	\label{glin}
	\eeq
	where $T(\vec{r})$ is the temperature profile obtained from the Joule heating equation, and
	$\gamma$ is the parameter characterizing the energy cost to create the inhomogeneous order parameter. This energy cost emerges as neighboring orders "bend" away from each other, and as such it is necessary to include in modelling spatially non-uniform systems \cite{Hohenberg_2015}. Fig. \ref{fig:dimer} plots the results of the MC simulations using the GL free energy given in Eq. \ref{glin}. We observe that the metallic domain of Nb-Nb weakened dimers are clearly formed at a higher voltage but not at the lower voltage, which can be viewed as the direct presentations for 'on' and 'off' states. We notice that the larger $\gamma$ is, the harder the metallic domain can form. This observation can be understood as follows. Because the larger $\gamma$ means higher energy cost to create the metallic domain, it is energetically favorable for the system to have the homogeneous order parameter if the temperature variation is small. This extra effect due to $\gamma$ term makes the 'on' and 'off' states even more distinguishable, which can be verified by experiments. For example, measurements of the T-EXAFS at different reservoir temperatures $T_{res}$ and different bias voltage can reveal the change of the Nb-Nb dimer length discussed above.
	
	\section{Discussion}
	\label{sec:diss}
	{\it Effects of disorder or impurity} --
	While our scenario is mainly based on the system with ideal crystalline structures, it is important to discuss qualitatively effects of disorder (or impurity). It is known that in the amorphous NbO$_x$ devices, several different crystals like NbO or Nb$_2$O$_{5-x}$ could be formed. We will not discuss the mixture of these radically different structures here, and we will focus on the effects of disorder or impurity on the system still in the NbO$_{2-x}$ structure. Generally speaking, the disorder could reduce the electrical conductivity because of the wavefunction localization. \cite{lee1985,singh2021} At higher temperature, the electrical conductivity could be recovered due to the phonon-assisted hoppings between localized states. However, the disorder results in stronger local electron-phonon coupling as well, which could enhance the local Joule heating effect, which is known to dictate the formation of robust conductive filaments in some amorphous samples. \cite{nano12050813} In other words, although the disorder and impurities could in general decrease the electrical conductivity, the Joule heating effect would not simply get smaller as implied by the simple term of $\sigma(T)\left(-\vec{\nabla}\phi\right)^2$. The Joule heating effect would be even stronger due to the quatum effect of wavefunction localization together with the stronger scattering of electrons. These effects are beyond the scope of our present model, and a more quantum mechanical approach like the Boltzmann equation developed by Allen and Liu is necessary to take the effects of disorder into account.\cite{allen2020} Nevertheless, these corrections will still make the temperature dependence of the electrical conductivity highly non-linear, thus all the qualitative features, including the crossover from three dimensions to one dimension and the existence of the threshold voltage, should remain unchanged even in the presence of disorder and impurities.
	
	{\it Multivalued IV curve} -- 
	The absence of the steady state solution in the insulating phase implies the possibility to obtain the multivalued current-voltage (IV) curve. The steady state solution is obtained by setting $\partial T/\partial t=0$. As a result, the existence of the steady state solution indicates that in the long time limit, the temperature distribution will reach a thermal equilibrium state regardless of the initial condition.
	On the other hand, the absence of the steady state solution implies that $\partial T/\partial t$ in Eq. \ref{jhfull} can not be ignored, and the temperature distribution would depend on the initial condition, even with the same boundary conditions. The dependence of the initial condition provides a natural way to have multiple solutions with a given voltage. This behavior in our model parallels the hallmark attribute of multi-level resistance states in memristor systems. For example, we argue that the existence of the multiple solutions is the source of the chaotic behavior observed in the previous work.\cite{kumar2017}
	
	\section{Conclusion}
	In this paper, we have presented a thermodynamic model for the resistivity switching in the crystalline NbO$_2$ rooted in the scenario of the MIT driven by a second order Peierls instability. Using the standard Ginzburg Landau theory, we have successfully reproduced the temperature dependence of the Nb-Nb dimer length in an excellent agreement with the previous measurement of temperature-dependent extended X-ray absorption fine structure spectroscopy (T-EXAFS). Moreover, by assuming the energy gap closing at the critical temperature $T_c$ due to the gradual weakening of the Nb-Nb dimerization, we have demonstrated that the electrical conductivity can be accurately fit by just one free parameter in a very wide range of the temperature ($1\leq10^3/T\leq 6$).
	
	Using the electrical conductivity and the thermal conductivity obtained either from accurate fit to experiments or from DFT calculations, we are able to solve the Joule heating equation to explore the temperature distribution in the system under the bias voltage. 
	From the analysis of the system in the geometry of a three-dimensional cylinder with axial symmetry,
	we have observed a crossover behavior from three dimensions to one dimension, which can be controlled by the ratio of the diameter to the longitudinal length ($2R/L_z$).
	We have found that due to the highly non-linear temperature dependence of the electrical conductivity, there exists a threshold voltage above which no steady state solution exists in the insulating phase.  
	We have further demonstrated that the threshold voltage can be engineered by $2R/L_z$, which can be used for tuning the desired memristive IV curve. Combining the temperature distribution obtained from the Joule heating equation with the Ginzburg Landau theory, we have simulated the evolution of metallic domains with the weakened Nb-Nb dimer, and our results can be further verified by the T-EXAFS measurement on the NbO$_2$ under bias voltage.
	
	Our work suggests a new route to achieve the voltage-induced resistivity switching in crystalline insulators. This concept could generally be applied to any materials that exhibit a second-order like metal-to-insulator transition. However, in order to have a scalable model that could describe NbO$_2$ based devices, several quantum mechanical effects related to the disorder, impurities, and amorphous structures have to be included. This extension is still in progress.
	
	\section{Acknowledgement}
	We thank W. A. Doolittle, S. A. Howard, C. N. Singh, M. Wahila, and B. White for valuable discussions.
	This work was supported by the Air Force Office of Scientific Research
	Multi-Disciplinary Research Initiative (MURI) entitled, “Cross-disciplinary
	Electronic-ionic Research Enabling Biologically Realistic Autonomous Learning
	(CEREBRAL)” under Award No. FA9550-18-1-0024 administered by Dr. Ali Sayir.
	S.W.O and W.-C.L. are grateful for the support of the summer faculty fellowship program (SFFP) sponsored by the Air-Force-Research-Lab (AFRL).
	
	\appendix 
	\section{Formalism of the Monte Carlo Simulation}
	\label{amc}
	
	With the GL free energy, the partition function can be constructed as
	\be
	{\mathcal Z} = \int d[M] e^{-\beta F(M)},
	\ee
	where $\beta = 1/k_B T$ and $k_B$ is the Boltzmann constant.
	We have employed the Monte Carlo (MC) simulation to compute the thermal average of the order parameter $M$ defined as
	\be
	\langle M\rangle_{th} = \frac{\int d[M] M e^{-\beta F(M)}}{{\mathcal Z}}
	\ee
	To simplify the MC simulation, it is helpful to introduce the dimensionless variable representing the temperature, $t\equiv T/T_c$, and $\beta F(M)$ can be rewriteen as
	\be
	\beta F(M) = \frac{1}{k_B T_ct}\left(\frac{a(T)}{2}M^2 + \frac{b(T)}{4}M^4 + \frac{c(T)}{6}M^6\right)
	\label{betaf}
	\ee
	The parameters in the GL free energy are chosen to be $a(T) = a_0 k_B(T-T_c)$, $b=b_0k_B(T-T_b)$, and $c(T) = k_B T_c$. $T_c$ is the critical temperature. $(a_0,b_0)$ are some dimensionless fitting parameters, and $T_b$ is another temperature scale capturing the non-linear effects at higher temperature. Substituting these fitting parameters into Eq. \ref{betaf}, we obtain
	\be
	\beta F(M) = \frac{1}{t}\left[\frac{a_0(t-1)}{2}M^2 + \frac{b_0(t-t_b)}{4}M^4 + \frac{1}{6}M^6\right],
	\label{betaf2}
	\ee
	where $t_b=T_b/T$ is a dimensionless parameter too. The advantage of using Eq. \ref{betaf2} in the MC simulation is that all the fitting parameters $(a_0,b_0,t_b)$ as well as the variables $(t.M)$ are dimensionless, which is a scale-invariant form suitable for studying systems with different physical sizes. 
	The paramteres used in this paper are $(a_0,b_0,t_b) = (1.0, 320.0, 0.6925)$.
	
	\section{First principles calculations of phonons and thermal conductivity}
	\label{adft}
	The calculations were performed using the density-functional theory (DFT) implemented in the Vienna Ab initio simulation package (VASP).\cite{kresse1993,kresse1996,kresse19962}  The local-density approximation (LDA)\cite{perdew1981} is used with a plane-wave cutoff 520 eV. The lattice parameters and atomic positions were relaxed with 8 x 8 x 12 Monkhorst–Pack electronic k-point mesh until all the forces acting on atoms were less than 0.01 mev/A°. The phonon dispersion and specific heat calculations were calculated using the Phonopy program\cite{togo2015} with supercell 2 x 2 x 2 and 4 x 4 x 6 Monkhorst–Pack electronic k-point mesh. The lattice thermal conductivity was calculated via the modified Debye–Callaway model proposed by Asen-Palmer using AICON code.\cite{fan2020}

\end{document}